\documentclass[floats,floatfix,showpacs,amssymb,prd,twocolumn,superscriptaddress,nofootinbib]{revtex4-1}

\usepackage{amssymb,amsmath,verbatim,mathtools,needspace,enumitem,etoolbox,graphicx,physics,microtype,afterpage,bm}

\usepackage[dvipsnames, usenames]{xcolor}

\definecolor{linkcolor}{rgb}{0.0,0.3,0.5}
\usepackage[unicode, colorlinks=true, linkcolor=linkcolor, citecolor=linkcolor, filecolor=linkcolor, urlcolor=linkcolor, linktocpage, breaklinks]{hyperref}
\usepackage[all]{hypcap}
\usepackage[T1]{fontenc}
\usepackage[utf8]{inputenc}
\usepackage{tabularx}
\usepackage{float}
\allowdisplaybreaks
\newcommand\prlsec[1]{\vspace{2mm}\noindent {\bf \emph{#1}}}

\def\be{\begin{equation}}
\def\ee{\end{equation}}
\newcommand{\beq}{\begin{eqnarray}}
\newcommand{\eeq}{\end{eqnarray}}

\begin{document}

\title{Spin-induced black-hole scalarization in Einstein-scalar-Gauss-Bonnet theory}

\author{Emanuele Berti}
\email{berti@jhu.edu}
\affiliation{Department of Physics and Astronomy, Johns Hopkins University, 3400 N. Charles Street, Baltimore, Maryland 21218, US}

\author{Lucas G. Collodel}
\email{lucas.gardai-collodel@uni-tuebingen.de}
\affiliation{Theoretical Astrophysics, Eberhard Karls University of T\"ubingen, D-72076 T\"ubingen, Germany}

\author{Burkhard Kleihaus}
\email{b.kleihaus@uni-oldenburg.de}
\affiliation{Institute of Physics, University of Oldenburg, D-26111 Oldenburg, Germany}

\author{Jutta Kunz}
\email{jutta.kunz@uni-oldenburg.de}
\affiliation{Institute of Physics, University of Oldenburg, D-26111 Oldenburg, Germany}

\date{\today}

 \begin{abstract} 
   We construct black hole solutions with spin-induced scalarization in a class of models where a scalar field is quadratically coupled to the topological Gauss-Bonnet term. Starting from the tachyonically unstable Kerr solutions, we obtain families of scalarized black holes such that the scalar field has either even or odd parity, and we investigate their domain of existence. The scalarized black holes can violate the Kerr rotation bound. We identify ``critical'' families of scalarized black hole solutions such that the expansion of the metric functions and of the scalar field at the horizon no longer allows for real coefficients. For the quadratic coupling considered here, solutions with spin-induced scalarization are entropically favored over Kerr solutions with the same mass and angular momentum.
 \end{abstract}

\maketitle

\prlsec{Introduction.}
Compact objects in gravity theories involving scalar degrees of freedom can undergo a phase transition induced by a tachyonic instability, known as ``spontaneous scalarization'': the solutions of the general relativistic field equations become unstable in certain regions of parameter space, developing scalar ``hair.'' This instability comes in different flavors.  Matter-induced spontaneous scalarization was originally proposed for compact neutron stars in scalar-tensor theories~\cite{Damour:1993hw}, but more recently it was shown that spontaneous scalarization is possible also in the absence of matter. Curvature-induced spontaneous scalarization of black holes (BHs) was first studied in Einstein-scalar-Gauss-Bonnet (EsGB) theories~\cite{Doneva:2017bvd,Silva:2017uqg,Antoniou:2017acq}. Charge-induced scalarization can also occur in Einstein-scalar-Maxwell theories~\cite{Herdeiro:2018wub}.

In this paper we consider EsGB theories with action
\begin{eqnarray}  
\label{act}
S=\frac{1}{16 \pi}\int d^4x \sqrt{-g} \left[R - \frac{1}{2}
 (\partial_\mu \phi)^2
 + f(\phi) R^2_{\rm GB}   \right],
\end{eqnarray} 
where we use geometrical units ($G=c=1$), and $\phi$ is a real scalar field coupled to the Gauss-Bonnet (GB) invariant
$R^2_{\rm GB} = R_{\mu\nu\rho\sigma} R^{\mu\nu\rho\sigma}- 4 R_{\mu\nu} R^{\mu\nu} + R^2$.

We will focus on the simple quadratic coupling function $f(\phi) = \frac{1}{8}\eta\phi^2$.
Early work showed that the GB invariant acts as a tachyonic mass term for the scalar (``curvature-induced'' BH scalarization) when $\eta>0$~\cite{Doneva:2017bvd,Silva:2017uqg}, but here we focus on the case $\eta<0$.  Recent work pointed out that the Kerr BH solutions of general relativity can still scalarize when $\eta<0$~\cite{Dima:2020yac}: the Gauss-Bonnet scalar for the Kerr metric can become negative close to the horizon, producing ``spin-induced'' scalarization when the dimensionless Kerr spin parameter $j\equiv J/M^2 \gtrsim 0.5$ ~\cite{Dima:2020yac}. This conclusion was confirmed analytically~\cite{Hod:2020jjy} and numerically~\cite{Doneva:2020nbb} by studying linear perturbations of Kerr BHs. These works concluded that the instability threshold depends on the Gauss-Bonnet (GB) coupling $\eta/M^2$ and on the (even or odd) symmetry of the scalar field under parity transformation.  For small values of the GB coupling and associated large values of $j$, the thresholds for even and odd parity differ, whereas for large values of the GB coupling the two thresholds almost coincide.

Here we show that stationary and axisymmetric BH solutions with spin-induced scalar hair do indeed exist in the nonlinear regime. We construct these BHs numerically, starting from the respective threshold solutions.  We then vary the input parameters to map out the domain of existence of scalarized BHs for both even- and odd-parity scalar fields. The expansions of the metric functions and of the scalar field at the horizon yield an analytic criterion to identify critical solutions that form the second boundary of the domain of existence.

We investigate the thermodynamical stability of these BH solutions by computing their entropy.  Solutions with curvature-induced scalarization are entropically disfavored with respect to Schwarzschild and Kerr BHs when $f(\phi)$ is quadratic~\cite{Silva:2017uqg,Collodel:2019kkx}, but they become entropically favored when we add a quartic term~\cite{Silva:2018qhn} or for exponential coupling functions~\cite{Doneva:2017bvd,Cunha:2019dwb}.  Linear perturbation theory shows that the entropically favored (disfavored) fundamental scalarized solutions are mode stable (unstable)~\cite{Blazquez-Salcedo:2018jnn,Silva:2018qhn,Macedo:2019sem}. Here we find that BH solutions with spin-induced scalarization are entropically favored over Kerr solutions with the same mass and angular momentum, but their dynamical stability remains an open question.

\prlsec{General framework.}
The generalized Einstein and scalar field equations follow by varying the action \eqref{act} with
respect to the metric $g_{\mu\nu}$ and the scalar field $\phi$:
\begin{equation}
G_{\mu\nu} =  T_{\mu\nu} \ , \ \ \ 
\nabla^2 \phi + \frac{df}{d\phi} R^2_{\rm GB}  =  0 \ , 
\label{eoms}
\end{equation}
where the effective stress-energy tensor
\begin{eqnarray}
T_{\mu\nu} &=&
-\frac{1}{4}g_{\mu\nu}\partial_\rho \phi \partial^\rho \phi 
+\frac{1}{2} \partial_\mu \phi \partial_\nu \phi \nonumber\\
&-&\frac{1}{2}\left(g_{\rho\mu}g_{\lambda\nu}+g_{\lambda\mu}g_{\rho\nu}\right)
\eta^{\kappa\lambda\alpha\beta}\tilde{R}^{\rho\gamma}_{\alpha\beta}\nabla_\gamma \partial_\kappa f(\phi) \ ,
\label{teff}
\end{eqnarray}
with $\tilde{R}^{\rho\gamma}_{\alpha\beta}=\eta^{\rho\gamma\sigma\tau}
R_{\sigma\tau\alpha\beta}$ and $\eta^{\rho\gamma\sigma\tau}= 
\epsilon^{\rho\gamma\sigma\tau}/\sqrt{-g}$.

To construct stationary, axially symmetric spacetimes with two commuting Killing vector fields ($\xi=\partial_t$ and $\eta=\partial_\varphi$) we employ a Lewis-Papapetrou--type ansatz~\cite{Wald:1984rg,Kleihaus:2000kg} \begin{eqnarray} \label{metric}
  ds^2&=&- b e^{F_0} dt^2 + e^{F_1} \left( d r^2+ r^2d\theta^2 \right) \nonumber\\
           &+&  e^{F_2} r^2\sin^2\theta (d\varphi+\omega dt)^2 ,
\end{eqnarray}
where $r$ is a quasi-isotropic radial coordinate, $r_{\rm H}$ is the isotropic horizon radius, $b=(1-\frac{r}{r_{\rm H}})^2$. The metric functions $F_i$ $(i=0,\,1,\,2$) and $\omega$ depend on the coordinates $r$ and $\theta$, and they are even under parity. The scalar field $\phi=\phi(r,\,\theta)$ can be either even or odd with respect to parity transformation, i.e.  $\phi_\pm(r,\pi-\theta)=\pm \phi_\pm(r,\theta)$.  Both parity-even and parity-odd scalar fields are consistent with the field equations, since the generalized Einstein equations are quadratic and the generalized Klein-Gordon equation is linear in $\phi$ (note that parity is a symmetry only when $f(\phi)$ is even in $\phi$).  Scalarized BHs with an even scalar field and no radial nodes are the fundamental scalarized solutions, whereas those with an odd scalar field are angularly excited solutions.

The proper set of boundary conditions is obtained by considering symmetry, regularity and asymptotic flatness of the solutions. This implies $F_i(\infty)=0 \quad (i=0,\,1,\,2),\,\quad \omega(\infty)=\phi(\infty)=0$ as $r\to \infty$.  For a massless scalar field one can construct an approximate solution of the field equations as a power series in $1/r$, with the dominant term being of monopole type in the even case and of dipole type in the odd case:
$\phi_{+}=Q/r+\dots$ and
$\phi_{-}=P \cos\theta/r^2+\dots$,
where $Q$ and $P$ are interpreted as the scalar charge and the dipole moment of the scalar field, respectively.

The boundary conditions at the event horizon, located at a surface of constant $r=r_{\rm H}$, are obtained by considering a power-series expansion in terms of $\delta=(r-r_{\rm H})/r_{\rm H}$:
$\partial_r F_0(r_{\rm H})=1/r_{\rm H}$,
$\partial_r F_1(r_{\rm H})=-2/r_{\rm H}$,
$\partial_r F_2(r_{\rm H})=-2/r_{\rm H}$,
$\omega(r_{\rm H})=\omega_{\rm H}$,
$\partial_r \phi(r_{\rm H})=0$,
where $\omega_{\rm H}$ is a constant.
On the symmetry axis ($\theta=0,\pi$), axial symmetry and regularity impose
$\partial_\theta F_i|_{\theta={0,\pi}} =0$ $(i=0,\,1,\,2)$,
$\partial_\theta \omega|_{\theta={0,\pi}} = \partial_\theta \phi|_{\theta={0,\pi}} =0$. 
Since all functions are either even or odd, it is sufficient to consider the range $0\leq \theta \leq \pi/2$ for the angular variable $\theta$ in the numerical calculations.  Consequently, we impose the following
 boundary conditions on the equatorial plane:
$\partial_\theta F_i|_{\theta=\pi/2} = 0$ $(i=0,\,1,\,2)$,
$\partial_\theta \omega|_{\theta=\pi/2} =
   \partial_\theta \phi_+|_{\theta=\pi/2} =
   \phi_-|_{\theta=\pi/2} =0$.

From the horizon metric we obtain the Hawking temperature~\cite{Wald:1984rg}
 \begin{eqnarray}
\label{TH}
T_{\rm H}=\frac{1}{2 \pi r_{\rm H}}e^{(F_0-F_1)/2}  \ .
 \end{eqnarray}
 In fact, the equation $G_{r}^\theta=T_{r}^\theta$ implies that $F_0/F_1$ (and therefore the Hawking temperature) is constant.  This observation can be used to test the numerical accuracy of our solutions.

\begin{figure}
\begin{center}
\includegraphics[width=.49\textwidth, angle =0]{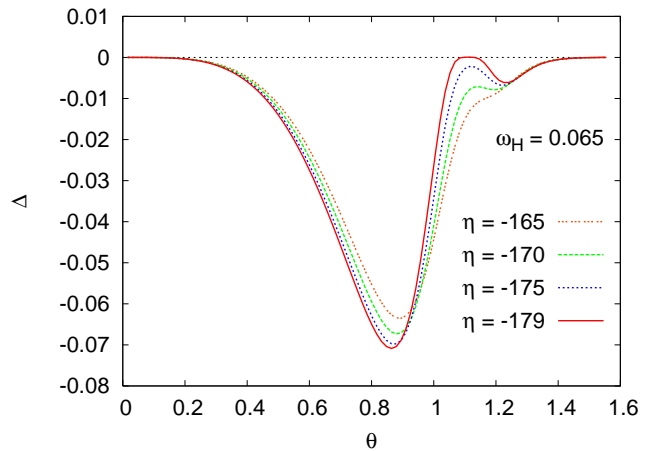}
\end{center}
\caption{Dimensionless discriminant as a function of $\theta$ for $\omega_{\rm H}=0.065$ and selected values of $\eta$. When $\eta< \eta_{\rm cr}\simeq -179$ the local maximum becomes positive, and scalarized BH solutions cease to exist. }
\label{fig_1n}
\end{figure}

\begin{figure*}
\begin{center}
\includegraphics[width=.49\textwidth, angle =0]{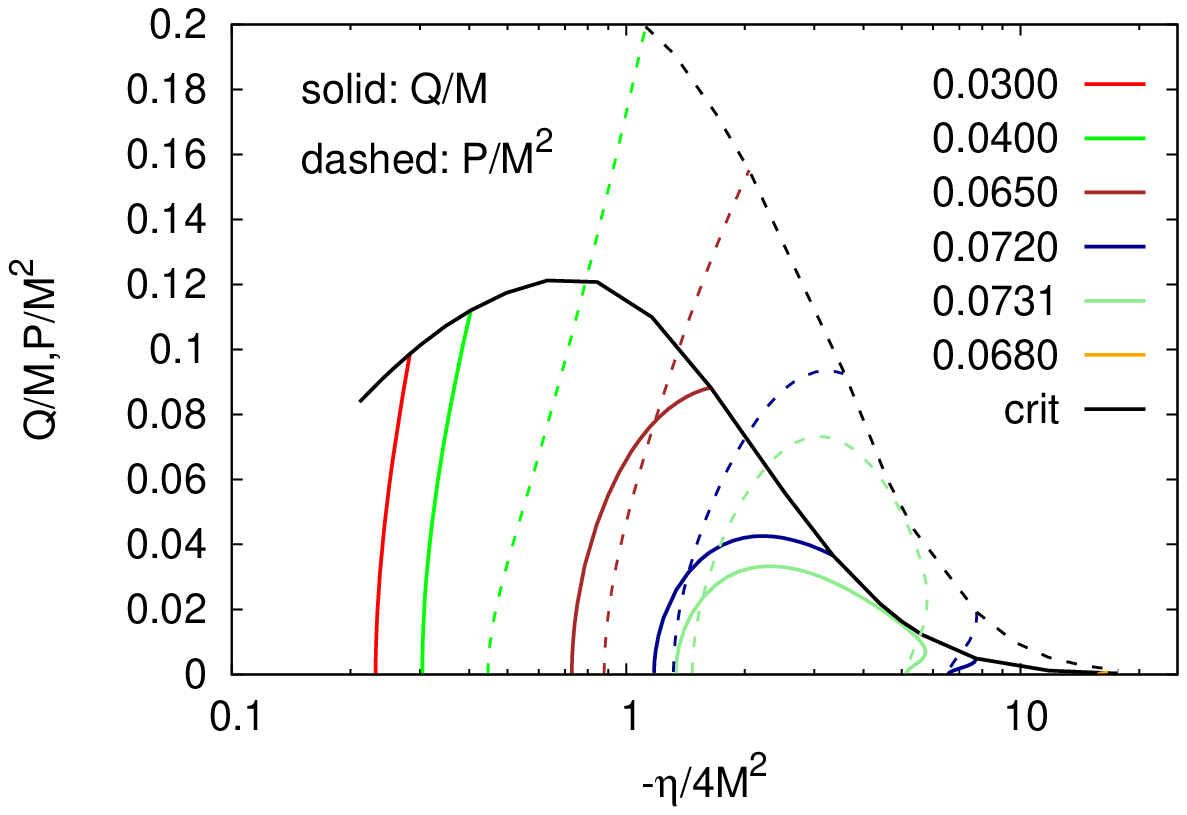}
\includegraphics[width=.49\textwidth, angle =0]{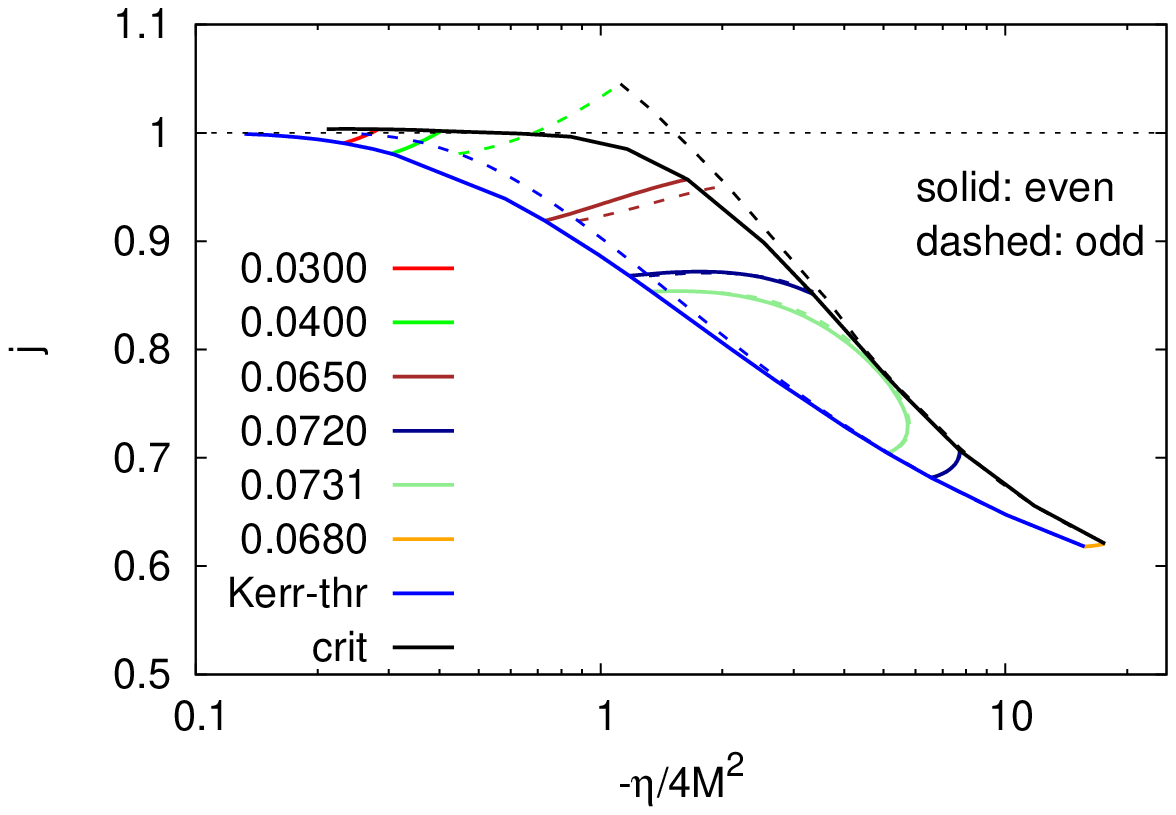}\\
\includegraphics[width=.49\textwidth, angle =0]{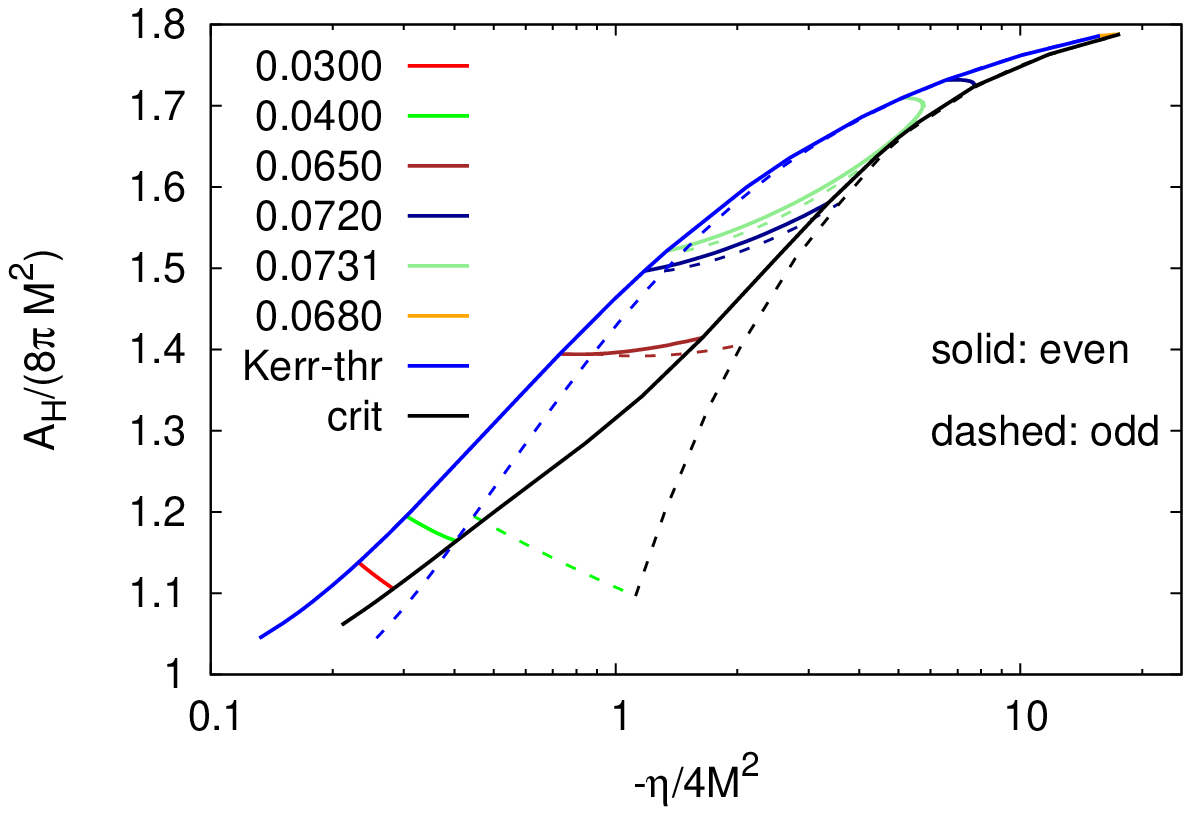}
\includegraphics[width=.49\textwidth, angle =0]{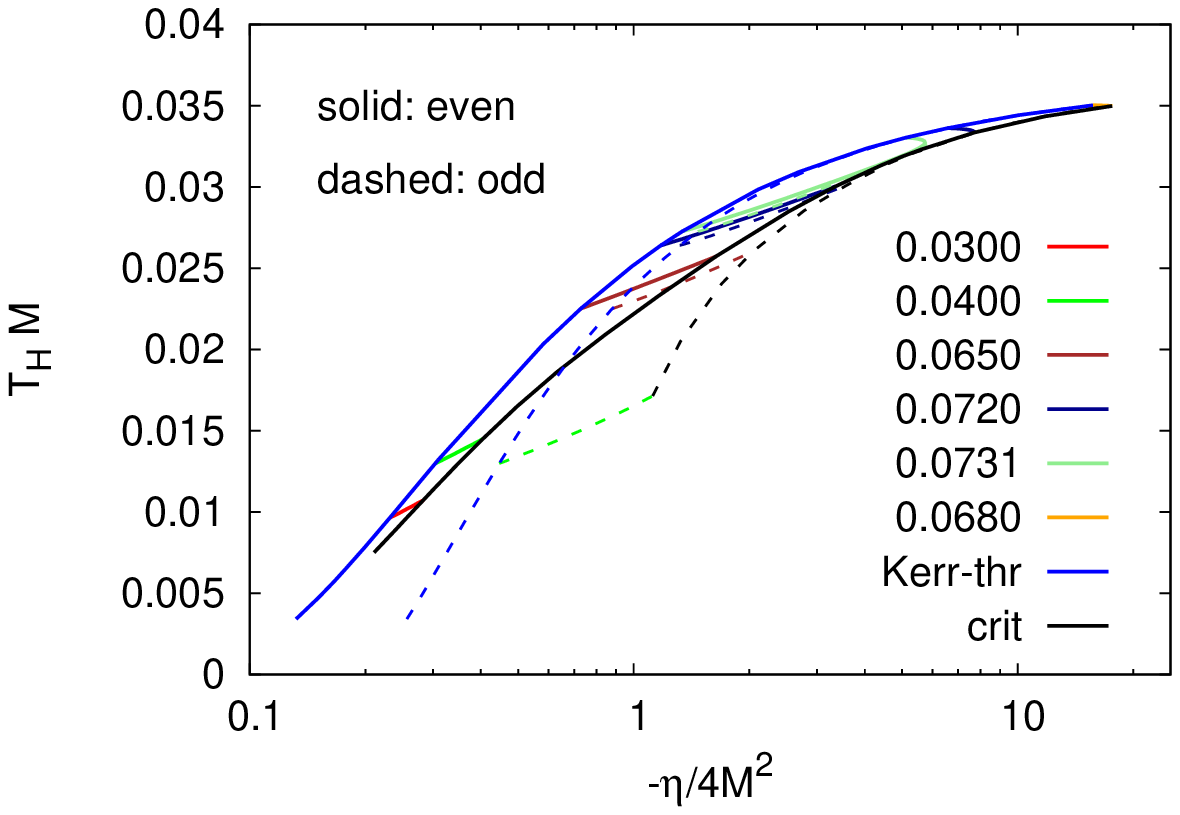}

\end{center}
\caption{Top left: dimensionless charge $Q/M$ of BHs with even scalar field (solid lines) and dimensionless dipole moment $P/M^2$ of BHs with odd scalar field (dashed lines). The other panels show the dimensionless angular momentum $j$ (top right), dimensionless horizon area $A_{\rm H}/8 \pi M^2$ (bottom left) and dimensionless Hawking temperature $T_{\rm H} M$ (bottom right) of even-parity BHs (solid lines) and odd-parity BHs (dashed lines). All quantities are shown for selected values of $\omega_{\rm H}$ as functions of $-\eta/4M^2$.
}
\label{fig_2n}
\end{figure*}

The horizon area is given by
 \begin{eqnarray}
\label{AH}
A_{\rm H}=2\pi r_{\rm H}^2\int_0^\pi d\theta 
\sin\theta e^{(F_1+F_2)/2}.
 \end{eqnarray}
The entropy of Kerr BHs is a quarter of the horizon area~\cite{Wald:1984rg}, but the entropy of EsGB BHs can be computed as an integral over the spatial cross section of the horizon~\cite{Wald:1993nt} and it acquires an extra contribution:
\begin{eqnarray}
\label{S-Noether} 
S=\frac{1}{4}\int_{\Sigma_{\rm H}} d^{2}x 
\sqrt{h}(1+ 2 f(\phi) \tilde R)\,,
\end{eqnarray} 
where $h$ is the determinant of the induced metric on the horizon and $\tilde R$ is the corresponding scalar curvature.

The mass $M$ and the angular momentum $J$ can be found from the asymptotic  behavior of the metric functions:
$g_{tt} %
  =-1+2M/r+\dots$,
$g_{\varphi t}%
=-2J \sin^2\theta/r+\nonumber \dots$.  

\prlsec{Numerical Results.}
To obtain the EsGB BHs with spin-induced scalarization we need to solve for the functions $(F_0, F_1, F_2, \omega;\, \phi)$ subject to the boundary conditions specified above, that guarantee regularity and asymptotic flatness.
{\sc REDUCE} files with the field equations and expansions are available from the authors upon request.

We provide three input parameters ($\eta$, $r_{\rm H}$ and $\omega_{\rm H}$) and we follow the numerical procedure of Refs.~\cite{Kleihaus:2000kg,Collodel:2019kkx}.
We introduce the radial variable $x=1-r_{\rm H}/r$, mapping the interval $[r_{\rm H},\infty)$ to the interval $[0,1]$, and discretize the equations on a nonequidistant grid in $x$ and $\theta$, covering the integration region $0\leq x \leq 1$ and $0\leq \theta \leq \pi/2$.
We perform the integrations using the package FIDISOL/CADSOL~\cite{Schoenauer:1989,Schauder:1992}, based on a Newton-Raphson method,
and we extract the physical properties of the BHs when the estimated truncation error is within the required accuracy, i.e., when the maximal numerical error for the functions at any point is estimated to be of order $10^{-3}$ or less.

We now address the second boundary of the domain of existence of scalarized BH solutions, given  by the set of critical solutions.
To this end, we consider higher-order terms of local solutions close to the horizon:
$F_0 = F_{0,{\rm H}} +\delta + f_{0,2}\delta^2/2 + \dots$,
$F_i = F_{i,{\rm H}} -2\delta + f_{i,2}\delta^2/2 + \dots$, $(i=1,\,2)$\,,
$\omega = \omega_{\rm H} + \omega_{2}\delta^2/2 + \omega_{3}\delta^3/6 +  \dots$,
$\phi = \phi_{\rm H} + \phi_{2}\delta^2/2 + \dots$.

We obtain equations for the coefficients of the higher-order terms $f_{0,2}$, $f_{1,2}$, $f_{2,2}$, $\omega_{3}$, and $\phi_{2}$ which allow us to express the higher-order coefficients in terms of $F_{0,{\rm H}}$, $F_{1,{\rm H}}$, $F_{2,{\rm H}}$, $\omega_{2}$, $\phi_{\rm H}$, and their first and second derivatives with respect to $\theta$. Solving these equations yields a quartic equation for $\phi_2$. The existence of real solutions of the quartic equation depends on the sign of the discriminant $D = \left(p/3\right)^3+\left(q/2\right)^2$ of the reduced cubic resolvent, $ v^3 + p v + q=0$. Real solutions exist if $D(\theta) \leq 0$ for $0\leq\theta\leq \pi/2$.  The numerical calculations show that $D(\theta)$ is always negative for $\omega_{\rm H}> 0.073$.  However, for $\omega_{\rm H}\leq 0.073$ the function $D(\theta)$ developes a local maximum, which tends to zero when $\eta$ is decreased to some critical value $\eta_{\rm cr}$. Solutions for $\eta <\eta_{\rm cr}$ cease to exist.

This is demonstrated in Fig.~\ref{fig_1n}, where we show the dimensionless discriminant
$\Delta \equiv\frac{4}{27} \frac{p^3}{q^2}+1$
for $\omega_{\rm H}=0.065$ and decreasing values of $\eta$.
Note however that for symmetric BHs with large angular momentum
the condition for real solutions is violated at the equator of the horizon.

We are now ready to discuss the physical properties of these spin-induced spontaneously scalarized black holes.
To map out their domain of existence, we have calculated numerous families of scalarized BH solutions with fixed horizon angular velocity $\omega_{\rm H}$ while varying the coupling constant $\eta$.

In Fig.~\ref{fig_2n} we show various BH properties as functions of the dimensionless coupling parameter $\eta/4M^2$ for families of solutions with fixed values of $\omega_{\rm H}$.  The different panels show the dimensionless scalar charge $Q/M$ (dipole moment $P/M^2$) for the fundamental even (odd) solutions (top left); the dimensionless angular momentum $j$ (top right); the dimensionless horizon area $A_{\rm H}/8 \pi M^2$ (bottom left); and the dimensionless Hawking temperature $T_{\rm H} M$ (bottom right).

Figure~\ref{fig_2n} provides important new insight into the domain of existence of BHs with spin-induced scalarization.  Bifurcation from the Kerr solutions takes place at some threshold solutions (``Kerr-thr'' in the legend) representing the first boundary of the domain of existence.  These thresholds are rather close for even and odd solutions, especially for large values of $|\eta/M^2|$.  The second boundary is given by the critical solutions (``crit'' in the legend) such that the discriminant $D(\theta)$ vanishes somewhere.  For large values of $|\eta/M^2|$ the threshold lines and the critical lines approach each other, and the domain of existence becomes narrower.

If present, the third boundary of the domain of existence should correspond to extremal scalarized BHs, which are numerically difficult to explore. The bottom-right panel of Fig.~\ref{fig_2n} shows that the Hawking temperature approaches zero in this limit.  The previously studied case of rotating dilatonic GB BHs suggests that these extremal scalarized solutions might not be regular~\cite{Kleihaus:2011tg,Kleihaus:2015aje,Cunha:2016wzk}.  Some of these solutions have angular momentum exceeding the Kerr bound $j=1$. In fact, the bound is already exceeded by nonextremal odd solutions when $|\eta/M^2| <1.53$.  For even solutions, the Kerr bound is exceeded only marginally when $|\eta/M^2| <0.55$.

\begin{figure}
\begin{center}
\includegraphics[width=.49\textwidth, angle =0]{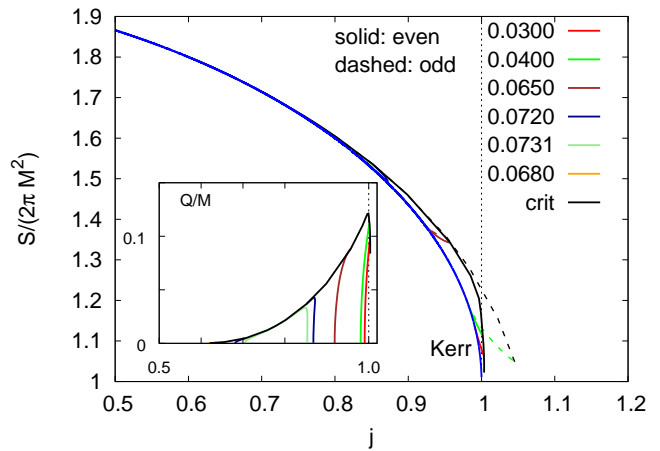}
\end{center} \caption{Dimensionless entropy $S/(2 \pi M^2)$ as a function of $j$ for odd- and even-parity BHs for selected values of $\omega_{\rm H}$. Inset: $Q/M$ as a function of $j$ for even-parity BHs.}
\label{fig_3n}
\end{figure}

The violation of the Kerr bound is also clear from Fig.~\ref{fig_3n}, where we plot the dimensionless entropy $S/2 \pi M^2$ as a function of $j=J/M^2$ for the same families of solutions. The inset of the figure shows the charge-to-mass ratio $Q/M$ as a function of $j$. Interpolation  yields a maximum value $Q/M=0.1225$ at $j=0.9989<1$. Most importantly, Fig.~\ref{fig_3n} allows us to draw a crucial conclusion: for a given mass and angular momentum, BHs with spin-induced scalarization have larger entropy than Kerr BHs, and therefore they are entropically favored.  Close to the Kerr bound, the area of even- and odd-parity BHs with spin-induced scalarization can exceed the area of their Kerr counterparts by about 30\%. This could have interesting observational consequences, e.g. in terms of telling them apart from Kerr BHs with very-long baseline interferometry of their shadow~\cite{Akiyama:2019cqa}.

\prlsec{Conclusions.}
Starting from even- and odd-parity threshold solutions, we have mapped out the domain of existence of BHs with spin-induced scalarization in EsGB theories with a quadratic coupling function.  The second boundary of the domain of existence corresponds to critical solutions beyond which the horizon expansion of the metric functions and of the scalar field no longer admit real coefficients.  If present, a third boundary should correspond to extremal scalarized BHs, but this regime is hard to explore numerically.
Scalarized BHs can violate the Kerr bound when $|\eta/M^2| <1.53$ ($|\eta/M^2| <0.55$) for odd (even) solutions. This violation seems to occur only in the vicinity of the extremal solutions, and it is of the order of 5\% (0.5\%) for odd (even) solutions.

Scalarized BHs are entropically favored over Kerr BHs with the same mass and angular momentum. If previous studies of curvature-induced scalarization are a useful analogy, this would suggest that BHs with spin-induced scalarization are (linearly) mode stable under perturbations. This may come as a surprise, since we have employed a simple quadratic coupling function; however we chose a negative coupling constant, in contrast with previous work on curvature-induced scalarization. The dynamical stability of BHs with spin-induced scalarization is an important open question that will require further work. Perturbations of rotating BHs in modified gravity are a notoriously difficult technical problem, because the equations are nonseparable (see e.g. Ref.~\cite{Cano:2020cao} for recent progress on scalar perturbations in a slow-rotation expansion). Time evolutions may provide a practical way to find out if these solutions are dynamically stable.

Last but not least, the problems of well-posedness, gravitational collapse, and gravitational waveforms from binary BH mergers in EsGB theories are very active research areas in analytical and numerical relativity~\cite{Witek:2018dmd,Ripley:2019hxt,Ripley:2019irj,Julie:2019sab,Ripley:2020vpk,Okounkova:2020rqw,Julie:2020vov,Witek:2020uzz}. The new solutions discussed in this paper may have important implications in this context. 

\prlsec{Acknowledgments.}
We thank C.~Herdeiro, E.~Radu, H.~O.~Silva, T.~P.~Sotiriou and N.~Yunes for sharing with us, while this paper was nearing completion, a manuscript in which they derive independently similar results for a different coupling function.
The authors gratefully acknowledge support by the
DFG Research Training Group 1620  \textit{Models of Gravity}
and the COST Actions CA15117 \textit{CANTATA} and CA16104 \textit{GWverse}. 
L.C. is thankful for the financial support obtained through the DFG Emmy Noether Research Group  
under  grant  no.   DO  1771/1-1.
E.B. is supported by NSF Grants No. PHY-1912550 and AST-2006538, NASA ATP Grants No. 17-ATP17-0225 and 19-ATP19-0051, and NSF-XSEDE Grant No. PHY-090003. This work has received funding from the European Union’s Horizon 2020 research and innovation programme under the Marie Skłodowska-Curie grant agreement No. 690904.

\bibliography{refs}

\end{document}